\documentclass[aoas,MSNbibl,nameyear,noautosecdot,dvips]{arximspdf}
\usepackage{algorithm}
\usepackage{graphicx}

\doi{10.1214/11-AOAS460}
\volume{5}
\issue{3}
\pubyear{2011}
\firstpage{2131}
\lastpage{2149}

\begin{document}
\begin{frontmatter}

\title{A method for visual identification of small sample subgroups and potential biomarkers}
\runtitle{Bivisualization with CUMBIA}

\begin{aug}
\author[A]{\fnms{Charlotte} \snm{Soneson}\corref{}\ead[label=e1]{lottas@maths.lth.se}}
\and
\author[A]{\fnms{Magnus} \snm{Fontes}\ead[label=e2]{fontes@maths.lth.se}}
\runauthor{C. Soneson and M. Fontes}
\affiliation{Lund University}
\address[A]{Centre for Mathematical Sciences\\
Lund University\\
S-221 00 Lund\\
Sweden\\
\printead{e1}\\
\hphantom{\textsc{E-mail:}\ }\printead*{e2}} 
\end{aug}

\received{\smonth{5} \syear{2010}}
\revised{\smonth{1} \syear{2011}}

%
\begin{abstract}
In order to find previously unknown subgroups in biomedical data and
generate testable hypotheses, visually guided exploratory analysis can
be of tremendous importance. In this paper we propose a new
dissimilarity measure that can be used within the Multidimensional
Scaling framework to obtain a joint low-dimensional representation of
both the samples and variables of a multivariate data set, thereby
providing an alternative to conventional biplots. In comparison with
biplots, the representations obtained by our approach are particularly
useful for exploratory analysis of data sets where there are small
groups of variables sharing unusually high or low values for a small
group of samples.
\end{abstract}

%

\end{frontmatter}

\section{\texorpdfstring{Introduction.}{Introduction}}
As the amount and variety of biomedical data increase, so does the hope
of finding biomarkers, that is, substances that can be used as
indicators of specific medical conditions. It can also be possible to
detect new, subtle disease subtypes and monitor disease progression. In
these latter cases an exploratory approach may be beneficial in order
to detect previously unknown patterns. Exploratory analysis methods
providing a visually representable result are particularly appealing
since they allow the unparalleled power of the human brain to be used
to find potentially interesting structures and patterns in the data.
The inability to interpret objects in more than three dimensions has
motivated the development of methods that create a low-dimensional
representation summarizing the main features of the observed data.
Probably the most well-known such method is Principal Components
Analysis (PCA) [\citet{Pearson01}; \citeauthor{Hotelling33a} (\citeyear{Hotelling33a,Hotelling33b})] which
provides the best approximation (measured by the Frobenius norm) of a
given rank to a data matrix, and which is used extensively [see, e.g.,
\citet{Alteretal00}; \citet{Rossetal03} for applications to gene expression
data]. One particularly appealing aspect of PCA is that its formulation
in terms of the singular value decomposition (SVD) provides also a
low-dimensional representation of the variables, which is directly
synchronized with the sample representation. This allows for a visually
guided interpretation of the impact of each variable on the patterns
seen among the samples. The joint visualization obtained by depicting
both the sample and variable representations in the same plot is
commonly referred to as a biplot [\citet{Gabriel71}; \citet{GowerHand96}].
Biplots have been used for visualization and interpretation of many
different types of data [e.g., \citet
{PhillipsMcNicol86}; \citet{DeCrespinDeBillyetal00}; \citet{Chapmanetal01}; \citet{Woutersetal03}; \citet{Parketal08}].

The usefulness of PCA is dependent upon the assumption that the
Euclidean distance between the variable profiles of a pair of samples
provides a good measure of the dissimilarity between the samples. It is
easy to imagine situations where this is not true, for example, if two
samples should be considered similar if they show similar, unusually
high or low values on only a small subset of the variables irrespective
of the values of the rest of the variables, or if the samples are
distributed along a nonlinear manifold. Furthermore, to be extracted by
the first few principal components, which are usually used for
visualization and interpretation, a pattern must encode a substantial
part of the variance in the data set. This means that small groups of
samples may be difficult to extract visually, even if they share a~characteristic variable profile.

To address the shortcomings of PCA and allow accurate visualization of
more complex sample configurations, a variety of generalizations and
alternatives to PCA have been proposed, such as projection pursuit
[\citet{FriedmanTukey74}; \citet{Huber85}], kernel PCA [\citet
{Scholkopfetal98}] and other manifold learning methods such as Isomap
[\citet{Tenenbaumetal00}], Locally Linear Embedding [\citet
{RoweisSaul00}] and Laplacian Eigenmaps [\citet{BelkinNiyogi03}]. Most
of these methods do not automatically provide a related variable
representation, which makes it more difficult to formulate hypotheses
concerning the relationship between the variables and the patterns seen
among the samples. In particular, this is true for methods based on
Multidimensional Scaling (MDS), which create a low-dimensional sample
representation based only on a given matrix of dissimilarities between
the samples.

In this paper we present CUMBIA (Computational Unsupervised Method for
BIvisualization Analysis), an exploratory MDS-based method for creating
a common low-dimensional representation of both the samples and the
variables of a data set. We use the term ``bivisualization'' to denote
both the process of creating low-dimensional sample and variable
visualizations and the resulting joint representations. When using
CUMBIA, we define a~measure of the dissimilarity between a sample and a
variable, and use this to calculate sample--sample and variable--variable
dissimilarities. All dissimilarities are put into a common
dissimilarity matrix. Finally, we apply classical MDS to obtain a joint
low-dimensional sample and variable representation. In this way, we
obtain a biplot-like result where the relations between samples and
variables can be readily explored. We apply CUMBIA to a synthetic data
set as well as real-world data sets, and show that it provides useful
bivisualizations which are often more informative than the biplots
obtained by conventional methods for data sets containing small sample
clusters sharing exceptional values for relatively few variables. In
many cases, PCA will fail to find these groups because they do not
encode enough of the variance in the data. We therefore believe that
the proposed method may be a valuable complement to existing methods
for hypothesis generation and visual exploratory analysis of
multivariate data sets.

\section{\texorpdfstring{Related work.}{Related work}}\label{relatedwork}
The approach described in this paper provides a joint visualization of
both samples and variables, which is particularly useful for data sets
containing small groups of samples sharing extreme values of few
variables. To our knowledge, this problem has not been specifically
addressed by previously proposed methods. In this section we compare
our approach to some existing methods for finding and visualizing
``interesting'' variable combinations and corresponding sample groups.

Constructing a biplot when the sample representation is obtained by PCA
is straightforward, as will be shown in Section \ref{biplots}. The \textit{nonlinear biplot} was introduced by
\citet{GowerHarding88} to
generalize this result to more general sample representations. For a
sample representation obtained by a~given ordination method, such as
PCA or MDS (based on a specific dissimilarity measure), Gower and
Harding construct the variable representation by letting one variable
at a time vary in a ``pseudo-sample,'' while keeping the values of the
other variables fixed at their mean values across the original samples.
Then, the (usually nonlinear) trajectory of the pseudo-sample in the
original sample representation is taken as a representation of the
variable. These trajectories can often be interpreted in much the same
way as ordinary coordinate axes. The approach described in our paper is
different from that in \citet{GowerHarding88}, since both samples and
variables are treated on an equal footing in the MDS and, hence, all
dissimilarities are used to obtain the low-dimensional representations.
Moreover, the nonlinear biplots may be hard to interpret when the
number of variables is large.

CUMBIA provides a joint low-dimensional representation of samples and
variables which highlights other patterns than conventional
multivariate visualization methods and where small groups of related
objects are often readily visible. Biclustering methods
[e.g., \citet{ChengChurch00}; \citet{Getzetal00}; \citet{Dhillon01}; \citet{Tanayetal02}; \citet{Wangetal02};
\citet{Ben-Doretal03}; \citet{Bergmannetal03}; \citet{MadeiraOliveira04}; \citet{BissonHussain08}; \citet{Regeetal08}; \citet{Leeetal10}] have
been proposed in different applications
with the explicit aim of extracting subsets of samples (documents) and
genes (words), so-called biclusters, such that the variables in a
subset are strongly related across the corresponding sample subset.
Some of the biclustering methods adopt a weighted bipartite graph
approach [\citet{Dhillon01}; \citet{Tanayetal02}]. Such an approach lies as the
foundation also for CUMBIA.
There are, however, important differences between biclustering methods
and CUMBIA. The genes in a bicluster are extracted to exhibit similar
profiles across the samples in the bicluster, while the variable
clusters found by CUMBIA are highly expressed in the closely related
samples compared to the rest. Furthermore, biclustering algorithms aim
to provide an exhaustive collection of significant biclusters, while
visualization methods like the one we propose provide a visual
representation of the most important features of the entire data set.
This representation immediately allows the researcher to find clusters,
detect outliers and obtain insights into the structure of the data
which can be used to generate hypotheses. A further potential advantage
of visualization methods compared to clustering is the ability to put
objects ``in between'' two clusters, and to visualize the relationship
between different clusters. In summary, although they are somewhat
similar, biclustering and CUMBIA have different objectives and
therefore are not likely to give the same results.

Projection pursuit methods [\citet{FriedmanTukey74}; \citet{Huber85}] are
designed to search for particularly ``interesting'' directions in a
multivariate data set, where ``interestingness'' can be defined, for
example, as multimodality or deviation from Gaussianity. PCA is one
example of a projection pursuit method, where the interesting
directions are those with maximal variance. In this special case, the
optimal directions can be obtained by solving an eigenvalue problem
but, in general, projection pursuit methods are iterative and the
result may depend on the initialization. If the projections onto the
extracted directions and the contributions of the variables to these
are visualized simultaneously, the result can be interpreted to some
extent like a biplot.

\section{\texorpdfstring{The CUMBIA algorithm.}{The CUMBIA algorithm}}
In the following, we let $X\in\mathbb{R}^{N\times p}$ denote a data
matrix, containing the measured values of $p$ random variables in $N$ samples.
We denote the element in the $i$th row and $j$th column of a matrix~$A$
by $A_{ij}$.
Furthermore, the Frobenius norm of an $m\times n$ matrix $A$ is defined~by
\[
\|A\|_F^2=\sum_{i=1}^m\sum_{j=1}^n|A_{ij}|^2.
\]

\subsection{\texorpdfstring{Biplots and the duality of the singular value
decomposition.}{Biplots and the duality of the singular value
decomposition}}\label{biplots}
In this section we will recapitulate how the singular value
decomposition allows us to represent both the samples and the variables
of a data set in lower-dimensional spaces. On a pair of such
low-dimensional spaces we can define a~bilinear real-valued function,
which when applied to a sample and a variable immediately approximates
the value for the variable in that sample. This bilinear function will
then be used to create a dissimilarity measure relating samples and variables.

The singular value decomposition (SVD) of a matrix $X\in\mathbb
{R}^{N\times p}$ with rank~$r$ is given by
\[
X=U\Lambda V^T,
\]
where $U=[u_1,\ldots ,u_r]\in\mathbb{R}^{N\times r}$, $V=[v_1,\ldots
,v_r]\in\mathbb{R}^{p\times r}$ and $\Lambda\in\mathbb{R}^{r\times r}$.
The columns of $U$ and $V$ are pairwise orthogonal and of unit length
(so $U^TU=V^TV=I_r$), and $\Lambda=\operatorname{diag}(\lambda_1,\ldots ,\lambda_r)$ is
a diagonal matrix containing the positive singular values of $X$ in
decreasing order along the diagonal. We will denote $U_s=[u_1,\ldots
,u_s]$, $V_s=[v_1,\ldots ,v_s]$, $\Lambda_s=\operatorname{diag}(\lambda_1,\ldots ,\lambda
_s)$ for $s\leq r$. The SVD can be used to create a rank-$s$
approximation of $X$ by
\[
X_s=U_s\Lambda_sV_s^T.
\]
We note that $X_r=X$. The Eckart--Young theorem [\citet{EckartYoung36}]
states that this approximation is optimal in the sense that
\[
\|X-X_s\|_F^2=\inf_{Y\in\mathbb{R}^{N\times p}|\operatorname{rank}(Y)=s}\|X-Y\|_F^2.
\]
The error in the approximation is given by
\[
\|X-X_s\|_F^2=\sum_{k=s+1}^r\lambda_k^2
\]
[\citet{EckartYoung36}].
Given a rank-$s$ approximation $X_s$ of a data matrix $X$, we want to
visualize its rows and columns in $s$-dimensional spaces (typically
$s=2$ or $3$). For a fixed $\alpha\in[0,1]$, we define $s$-dimensional
spaces~$\mathbf{V}_s$ and $\mathbf{U}_s$ as the span of the orthogonal
columns of $V_s\Lambda_s^{1-\alpha}$ and $U_s\Lambda_s^\alpha$,
respectively. Next, we rewrite $X_s$ as
\[
X_s=(U_s\Lambda_s^\alpha)(V_s\Lambda_s^{1-\alpha})^T.
\]
This shows that the rows of $U_s\Lambda_s^\alpha$ can be seen as the
coordinates for the approximated samples (the rows of $X_s$) in the
space $\mathbf{V}_s$. Similarly, the rows of $V_s\Lambda_s^{1-\alpha}$
can be seen as the coordinates for the approximated variables in the
space $\mathbf{U}_s$. Hence, we take the $N$ rows of $U_s\Lambda
_s^\alpha$ as the $s$-dimensional representations of the samples, and
the $p$ rows of $V_s\Lambda_s^{1-\alpha}$ as the $s$-dimensional
representations of the variables. Choosing $\alpha=1$ corresponds to
conventional PCA where the low-dimensional sample representation is
given by the rows of $U_s\Lambda_s$ and the \textit{principal components}
(PCs) are the columns of $V_s$ [\citet{CoxCox01}; \citet{Jolliffe02}]. With this
choice of $\alpha$, the PCA representation provides an approximation of
the Euclidean distances between the samples of the data set [\citet
{Jolliffe02}]. Choosing instead $\alpha=0$ would approximate the
Euclidean distances between the variables.

We next define bilinear functions $(\cdot,\cdot)_s\dvtx \mathbf{V}_s\times
\mathbf{U}_s\rightarrow\mathbb{R}$, by
%
\begin{equation}(\mathbf{a},\mathbf{b})_s:=\sum_{k=1}^sa_kb_k,\label{biplot}
\end{equation}
where $\{a_k\}_{k=1}^s$ and $\{b_k\}_{k=1}^s$ are the coordinate
sequences of $\mathbf{a}$ and $\mathbf{b}$ in $\mathbf{V}_s$ and
$\mathbf{U}_s$, respectively. We note that the value for variable
$\mathbf{w}_j$ in sample $\mathbf{s}_i$ can be computed as
%
\begin{equation}X_{ij}=\sum_{k=1}^r(U\Lambda^\alpha)_{ik}(V\Lambda
^{1-\alpha})_{jk}=(\mathbf{s}_i,\mathbf{w}_j)_r\label{exactvalue}
\end{equation}
and approximated by
%
\begin{equation}(X_s)_{ij}=\sum_{k=1}^s(U_s\Lambda_s^\alpha
)_{ik}(V_s\Lambda_s^{1-\alpha})_{jk}=(\mathbf{s}_i,\mathbf{w}_j)_s
\label
{approximatevalue}
\end{equation}
for $s\leq r$.

In classical biplots, the samples are represented by the rows of
$U_s\Lambda_s^\alpha$ and the variables are represented by the rows of
$V_s\Lambda_s^{1-\alpha}$ in the same low-dimensional plot [\citet
{CoxCox01}]. Then it follows from (\ref{biplot}) and (\ref
{approximatevalue}) that the value of the variable $\mathbf{w}_j$ in
the sample $\mathbf{s}_i$ can be approximated by taking the usual
scalar product between the coordinate sequences for $\mathbf{s}_i$ and
$\mathbf{w}_j$ [\citet{Gabriel71}]. This makes it possible to use the
low-dimensional biplots to visually draw conclusions about the
relationships between groups of samples and variables.

\subsection{\texorpdfstring{Creating a joint dissimilarity matrix for samples and variables.}{Creating a joint dissimilarity matrix for samples and variables}}
Using the value of $(\mathbf{s}_i,\mathbf{w}_j)_{s}$ as a measure of
the similarity between sample $\mathbf{s}_i$ and variable $\mathbf
{w}_j$, we define the squared dissimilarity between $\mathbf{s}_i$ and
$\mathbf{w}_j$ as
%
\begin{equation}d^2_s(\mathbf{s}_i,\mathbf{w}_j)=\lambda_1-(\mathbf
{s}_i,\mathbf{w}_j)_{s},\label{Distance}
\end{equation}
where $\lambda_1$ is the largest singular value of $X$ (this is a
natural choice, making all dissimilarities nonnegative). We note that
this is just one way of transforming a measure of similarity to a
dissimilarity, and that there could be other possible transformations.
To define the dissimilarities between two objects of the same type
(i.e., two samples or two variables), we create a weighted bipartite
graph. In this graph, each sample is connected to all variables, and
each variable to all samples. The weight of an edge is taken as the
dissimilarity between the corresponding nodes, calculated by (\ref
{Distance}). The dissimilarity~$d_s(\mathbf{s}_i,\mathbf{s}_j)$ between
two samples [or $d_s(\mathbf{w}_i,\mathbf{w}_j)$ between two variables]
is then defined as the shortest distance between the corresponding
nodes in the weighted graph. Together with (\ref{Distance}), this
yields a joint $(N+p)\times(N+p)$ dissimilarity matrix containing the
dissimilarities between all pairs of objects. In this work, we restrict
our attention to paths consisting of only two edges (i.e., going from
one sample to another via only one variable, and vice versa), which
will allow us to compute the sample--sample and variable--variable
dissimilarities without actually creating the graph. By allowing more
complex paths, two samples could be considered similar if they are both
similar to a~third sample, even if these similarities are due to
completely different sets of variables. However, this may not be
desirable in an application where the goal is to find biomarkers, since
these should ideally be expressed very strongly in all samples in the
corresponding group.\footnote{It could be useful, for example, in a
document classification application, where documents discussing the
same topic with different words may be considered similar since both
share words with a third document on the same topic [\citet{BissonHussain08}].}

From (\ref{exactvalue}), we note that if we choose $s=r$, the
dissimilarity between a~sample and a variable depends only on $\lambda
_1$ and the expression value of the variable in that sample. If we
choose $s<r$, (\ref{approximatevalue}) implies that the dissimilarity
$d_s(\mathbf{s}_i,\mathbf{w}_j)$ is calculated from the approximated
value of $X_{ij}$ obtained by SVD. Using $s<r$ may be an advantage from
a noise reduction point of view, since we in this case discard the
smallest singular values and represent the data matrix only by its
dominant features. It is important to note that by using a very small
value of $s$, we may discard a large part of the true signal as well.

\subsection{\texorpdfstring{Creating a low-dimensional representation of samples and variables.}{Creating a low-dimensional representation of samples and variables}}
To obtain a low-dimensional representation of the samples and variables
from the dissimilarity matrix $D$, we apply classical MDS [\citet
{Torgerson52}]. Classical MDS finds a low-dimensional projection with
interpoint Euclidean distances collected in the matrix $\tilde{D}$,
such that
\[
\|C(D)-C(\tilde{D})\|_F
\]
is minimized [\citet{Mardia78}; \citet{CoxCox01}]. Here,
\[
C(D)=-\tfrac{1}{2}JD^2J,
\]
where $(D^2)_{ij}=(D_{ij})^2$, $J=I_n-\frac{1}{n}\mathbf{1}\mathbf
{1}^T$ with $\mathbf{1}$ denoting the column vector with all entries
equal to one, and $n$ is the number of objects. The optimal
representation is obtained by the top eigenvectors of $C(D)$, scaled by
the square root of the corresponding eigenvalues. If $D$ is a Euclidean
distance matrix, $C(D)$ is a corresponding inner product matrix and
classical MDS returns the projections onto the principal components
[\citet{Gower66}]. If $D$ does not correspond to distances in a
Euclidean space, then $C(D)$ is not positive semidefinite and, hence,
some eigenvalues of $C(D)$ are negative [\citet{CoxCox01}]. In this case
it is common either to add a suitable constant to all off-diagonal
entries of $D$, thereby making it correspond to a~distance matrix in a
Euclidean space [\citet{Cailliez83}], or to simply ignore the negative
eigenvalues and compute the representation from the eigenvectors
corresponding to the largest positive eigenvalues. In this paper we
apply the latter approach.

Algorithm \ref{algorithm1} summarizes the main steps of CUMBIA and a
small schematic example is provided in the Supplementary Material.

\begin{algorithm}
\caption{CUMBIA}
\label{algorithm1}
Input: Data matrix $X\in\mathbb{R}^{N\times p}$, number of paths to
average over ($K$).
\begin{enumerate}[5.]
\item\label{compdiss} Compute the dissimilarities for all
sample--variable pairs using (\ref{Distance}).
\item Create a weighted bipartite graph, where the weight of an edge
between a sample and a variable is equal to the dissimilarity computed
in step \ref{compdiss}.
\item Compute the dissimilarities for all sample--sample and
variable--variable pairs as distances in the graph. Average over the $K$
shortest paths.
\item Collect all dissimilarities in a common dissimilarity matrix and
perform classical MDS.
\item Visualize the result in a few dimensions.
\end{enumerate}
\end{algorithm}

\section{\texorpdfstring{Practical considerations.}{Practical considerations}}
\subsection{When will a pair of objects be considered similar?}
From the construction of the dissimilarity (\ref{Distance}) between
samples and variables and the computation of sample--sample
dissimilarities as graph distances it follows that two samples are
considered similar if they share a high value for a single variable.
This means that the proposed dissimilarity measure emphasizes mainly
the large values in the data matrix $X$. Hence, as for PCA and many
other multivariate\vadjust{\goodbreak} techniques, the scale of the variables will
influence the results.
The data can be normalized to the same scale before these methods are
applied, for example, by subtracting the mean value and dividing by the
standard deviation of each variable to obtain a matrix of $z$-scores.

With the proposed dissimilarity measure, two identical samples will
almost certainly have a positive dissimilarity with each other, which
is somewhat counterintuitive. In this paper we put the dissimilarity
between identical samples or variables to zero but other solutions are
possible, such as multiplying the dissimilarity values with function
values which are zero for identical objects and rise steeply toward one
as the objects become more dissimilar. The function can be, for
example, a sigmoidal function of the Euclidean distance between the
objects. In many practical applications, identical or near-identical
objects are very uncommon and, therefore, this is not likely to have a
major impact on the results from real data sets.

It is important to note that from the construction of the
bivisualization, it follows that it should be interpreted in terms of
the relative distances between objects and not, as in conventional
principal components biplots, in terms of the inner products between
samples and variables.

\subsection{\texorpdfstring{Computational considerations.}{Computational considerations}}
Creating a graph with edges connecting every sample--variable pair and
computing the distances in the graph can be a time-consuming task if
the number of variables or samples is large. However, by the
construction of the dissimilarity measure (\ref{Distance}), the
dissimilarity matrix can be computed directly from the matrix $X_s$ and
the largest singular value of $X$ by
%
\begin{eqnarray}\label{compdistance}
 d_s(\mathbf{s}_i,\mathbf{s}_j) &=& \min_{1\leq k\leq p}
\bigl(\sqrt{\lambda_1-(X_s^T)_{ki}}+\sqrt{\lambda_1-(X_s^T)_{kj}}
\bigr), \nonumber\\
\eqntext{1\leq i,j\leq N,\ \mathbf{s}_i\neq\mathbf{s}_j,}
\\
d_s(\mathbf{w}_i,\mathbf{w}_j) &=& \min_{1\leq k\leq N} \bigl(\sqrt
{\lambda_1-(X_s)_{ki}}+\sqrt{\lambda_1-(X_s)_{kj}} \bigr),\nonumber\\\\[-20pt]
 \eqntext{1\leq i,j\leq p,\ \mathbf{w}_i\neq\mathbf{w}_j,} \\
 d_s(\mathbf{s}_i,\mathbf{w}_j) &=& \sqrt{\lambda_1 - (X_s)_{ij}},
  \qquad 1\leq i\leq N,\ 1\leq j\leq p,
\nonumber
\end{eqnarray}
where we let $(X_s)_{ki}$ denote the element in the $k$th row and $i$th
column of $X_s$, and similarly for $X_s^T$.
The self-dissimilarities are always put to zero.
However, also the classical MDS has a high computational complexity,
which implies that the number of samples and variables should not be
too large. Hence, in large data sets such as genome-wide expression
data sets a variable selection should be performed before applying
CUMBIA. The variable selection can be guided by expert knowledge in the
field. Alternatively, the algorithm can initially be applied, for
example, to the probes from each chromosome individually or to random
subsets of the variables.

\subsection{\texorpdfstring{Stability to outliers.}{Stability to outliers}}
Since the visualization algorithm as described above depends only on
the shortest path between two objects in the graph, it is sensitive to
outliers, for example, large measurement errors for single variables.
The stability can be increased by averaging over the $K$ shortest paths
between any pair of samples (or variables), but it should be noted that
choosing a large $K$ decreases the ability to detect very small sample
and variable groups. Such a stabilization also permits a
computationally efficient implementation, by replacing the $\min$ value
in (\ref{compdistance}) by the average of the $K$ smallest values. It
is possible to choose different values of $K$ for sample pairs and
variable pairs.

\subsection{\texorpdfstring{Emphasizing both over- and underexpressed variables.}{Emphasizing both over- and underexpressed variables}}
As described above, CUMBIA emphasizes the variables which are
overexpressed in a~group of samples, and these variables and samples
are placed close to each other in the low-dimensional joint
visualization. However, the dissimilarities between jointly
underexpressed variables are also calculated based on their highest
expression values. Since these may be very low, a group of variables
which are jointly underexpressed may obtain large dissimilarities with
each other. This means that these variables may not form a tight
cluster located far from the corresponding samples, as in PCA. The
method can be adjusted to emphasize also this type of relationship, by
changing the calculation of the sample--sample and variable--variable
dissimilarities (see the Supplementary Material for details).

\section{\texorpdfstring{Applications.}{Applications}}
In order to illustrate and visually evaluate the characteristics of
CUMBIA, we apply it to synthetic data as well as real-world data sets
and compare the results to other methods. The first two examples
illustrate the benefits of using CUMBIA for visualization of data sets
where the nonrandom variation is attributable to a small group of
variables being overexpressed in few samples, and the third example
shows that CUMBIA performs well also in an example where the
informative features encode a~large part of the variance in the data
set, which is the situation where PCA is most useful. Taken together,
these examples suggest that CUMBIA can provide useful visualizations in
many different situations and since the feature extraction is not
guided by variance content, we can obtain other insights into the data
structure than with, for example, PCA. In all examples, we compute the
dissimilarity between pairs of samples (or pairs of variables) by
averaging over the $K=3$ shortest paths in the graph. We use the
original formulation of the algorithm, which means that we will focus
on finding overexpressed variables. Furthermore, we use $s=r=\operatorname{rank}(X)$
to calculate the CUMBIA dissimilarity matrix (\ref{compdistance}), that
is, we apply the method to the values in the original data matrix.

We compare the visualizations obtained by CUMBIA to the biplots
obtained from PCA as well as results from a projection pursuit
algorithm and the SAMBA biclustering method [\citet{Tanayetal02}]. We
applied the projection pursuit method implemented in the FastICA
package (version 1.1-11) [\citet{HyvarinenOja00}] for R. This method
searches for directions where the data show the largest deviation from
Gaussianity. First, the data are whitened by projecting onto the
leading $d$ principal components, and then the projection pursuit
directions are sequentially extracted from the whitened data. Since
these directions are not naturally ordered, we show all $d$ projection
pursuit components and the corresponding sample representations in the
Supplementary Figures. SAMBA was applied through the EXPANDER software
(version 5.09) [\citet{Shamiretal05}].
As noted in Section~\ref{relatedwork}, the aim of the biclustering
methods is slightly different than that of CUMBIA, and the comparison
mainly serves as an illustration of the different knowledge that can be
visually extracted using CUMBIA compared to these methods. More
examples showing the effect of choosing different parameter values in
CUMBIA are available in the Supplementary Material [\citet
{SonesonFontessupplement}].

\subsection{\texorpdfstring{Synthetic data set.}{Synthetic data set}}

\begin{figure}[t!]

\includegraphics{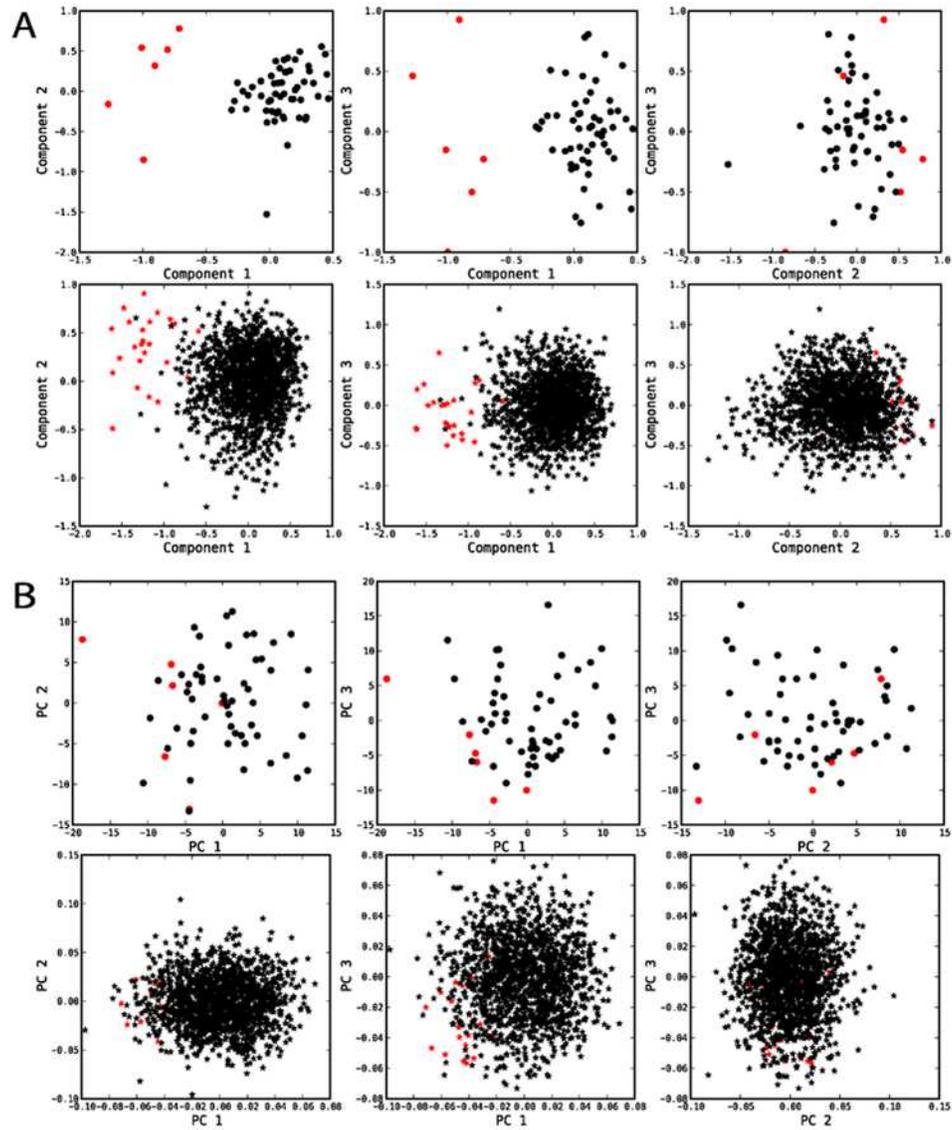}

\caption{Low-dimensional representation of samples and variables from
the synthetic data set, obtained by CUMBIA (panel \textup{A}) and PCA (panel \textup{B}).
Sample representations are shown in the top row, and corresponding
variable representations are shown below. Each subfigure shows the
representation with respect to two of the three first components. Red
markers represent the six samples and 25 variables which are simulated
to be closely related. Black markers represent the other 54 samples and
1,475 variables in the data set (PC---principal component).}
\label{simex1}
\end{figure}

We simulate a data matrix $X$ consisting of 60 samples and 1,500
variables by letting
\[
x_{ij}\in
\cases{\displaystyle
\mathcal{N}(2,1),&\quad$1\leq i\leq6$, $1\leq j\leq25$,\cr\displaystyle
\mathcal{N}(0,1),&\quad otherwise.
}
\]
Hence, there is a small group of 25 variables characterizing a group of
six samples. Each variable is mean-centered and scaled to unit variance
across all samples.
Figure~\ref{simex1} shows the low-dimensional representations of
samples and variables obtained by CUMBIA and PCA. We note that the
small size of the related sample and variable group makes it impossible
to extract clearly with PCA in the first three components. Even if more
components are included, the two groups do not separate (data not
shown). We use $d=10$ principal components to whiten the data before
applying the projection pursuit algorithm. The small group of six
samples is not visible in any of the projection pursuit components
either (see the Supplementary Figures). In contrast, the first CUMBIA
component discriminates the small sample group and the related
variables from the rest. Scree plots for CUMBIA and PCA are available
in the Supplementary Material. Applying SAMBA to the synthetic data set
does not return any biclusters.

%

\subsection{\texorpdfstring{Microarray data set---human cell cultures.}{Microarray data set---human cell cultures}}

Next, we consider a real microarray data set, from a study of gene
expression profiles from 61 normal human cell cultures. The cell
cultures are taken from five cell types in 23 different tissues or
organs, in total 31 different tissue/cell type combinations. The data
set was downloaded from the National Center for Biotechnology
Informations (NCBI) Gene Expression Omnibus (GEO,
\href{http://www.ncbi.nlm.nih.gov/geo/}{http://www.ncbi.}
\href{http://www.ncbi.nlm.nih.gov/geo/}{nlm.nih.gov/geo/}, data set GDS1402).
The original data set consists of 19,664 variables. We remove the
variables containing missing values (2,741 variables) or negative
expression values (another 517 variables), and the remaining values are
$\log_2$-transformed.

\begin{figure}

\includegraphics{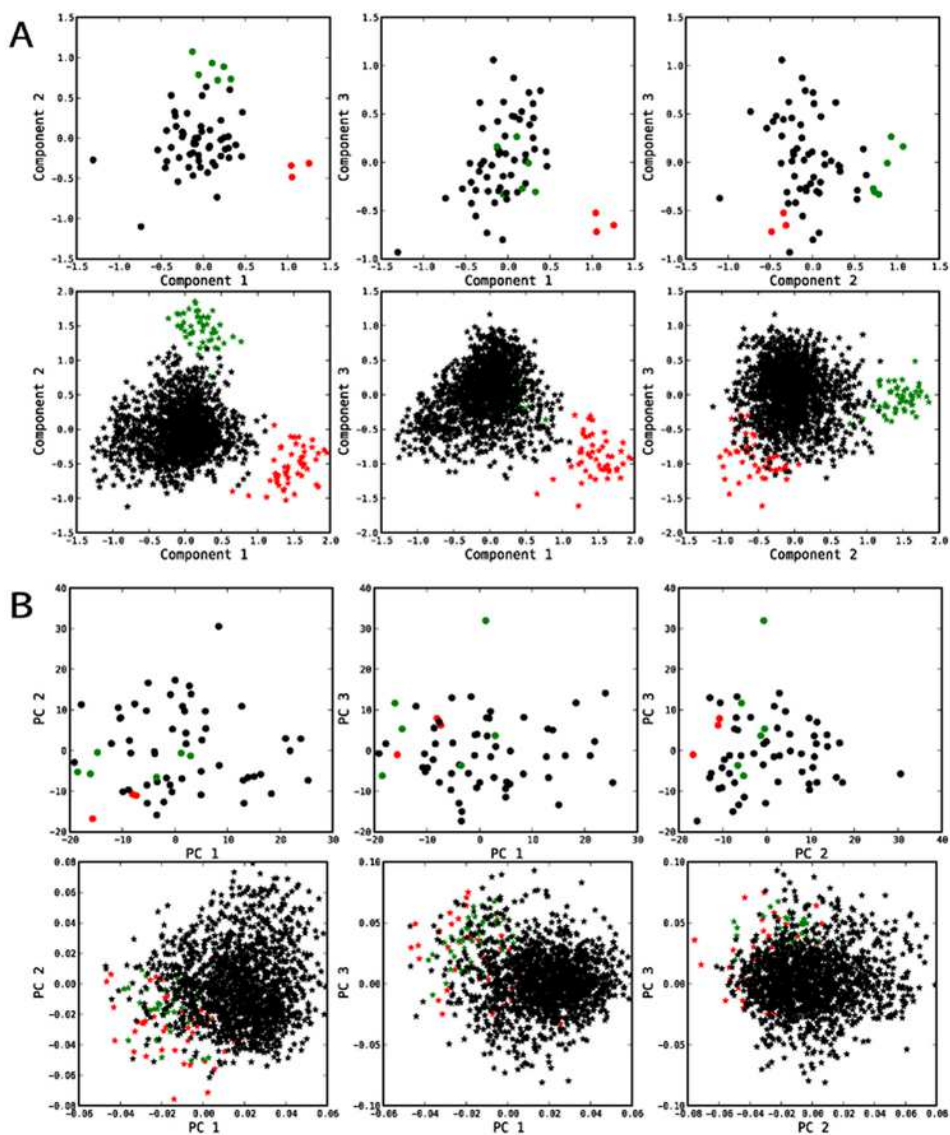}

\caption{Low-dimensional representation of samples and variables from
the human cell culture microarray data set, obtained by CUMBIA (panel
\textup{A}) and PCA (panel \textup{B}). Red markers represent samples from the cardiac
stromal cells ($N_1=3$), and the 50 variables with highest
discriminative power for this sample group, respectively. Green markers
similarly represent the umbilical artery endothelial cells ($N_2=6$)
and the corresponding variables. Black markers represent samples from
all other subgroups, and the 1,500 variables from the original data set
which are least discriminating in an $F$-test contrasting all 31
tissue/cell type combinations in the data set.}
\label{realex1}
\end{figure}

To illustrate the ability of CUMBIA to detect small sample and variable
clusters, we create a new data set from a subset of the variables in
the microarray data set.
We select two of the nontrivial sample subgroups, cardiac stromal cells
($N_1=3$) and umbilical artery endothelial cells ($N_2=6$). For each of
these sample subgroups and for each variable, we perform a $t$-test
contrasting the selected subgroup against all other samples. For each
of the two subgroups, we include the 50 variables having the highest
positive value of the $t$-statistic. We further extend the new data set
with the 1,500 variables showing the least discriminative power (the
lowest value of the $F$-statistic) in an $F$-test contrasting all 31
subgroups. Finally, all variables are mean centered and scaled to unit
variance across the samples. The final data set now consists of
$p=1\mbox{,}600$ variables and $N=61$ samples. This data set contains two
relatively small sample groups, each of which is characterized by high
values for a small subset of the variables. Furthermore, the vast
majority ($93.75\%$) of the variables are not related to any of the
predefined subgroups. Figure~\ref{realex1} shows the low-dimensional
representations of the samples and variables obtained by CUMBIA (panel
A) and PCA (panel~B).
The first two CUMBIA components successfully pick up the two small
sample subgroups as well as the variables which are responsible for
their close relation. These patterns do not encode enough variance to
be seen in any of the three first principal components (panel~B). In
the projection onto the fourth and fifth principal components, the
three cardiac stromal cell samples are visible as well as four of the
six umbilical artery endothelial cells (data not shown). Clearly, by
considering not only the variance of the extracted components as a
measure of informativeness, CUMBIA highlights other features than PCA.
Scree plots are available as the Supplementary Material. We used $d=10$
principal components for the whitening preceding the projection pursuit
algorithm, which is able to detect the group of cardiac stromal cells,
but the umbilical artery cells are considerably harder to extract (see the
Supplementary Figures). The projection pursuit algorithm further finds
one single umbilical artery cell occupying one component together with
a group of underexpressed variables. By modifying the CUMBIA algorithm
to search for both over- and underexpressed variables, we also find
this pattern (see the Supplementary Material, Figure S2). For this data
set, SAMBA returns 26 biclusters with significant overlaps. Eleven of
these contain two of the cardiac stromal cells (but none of them
contain all three). Eight biclusters contain at least two umbilical
artery endothelial cell samples (one contains all six). Again, we note
that the purpose of biclustering is not quite the same as the purpose
of visualization which can also be seen in this example.

\subsection{\texorpdfstring{MicroRNA data set---leukemia cell lines.}{MicroRNA data set---leukemia cell lines}}

\begin{figure}[t!]

\includegraphics{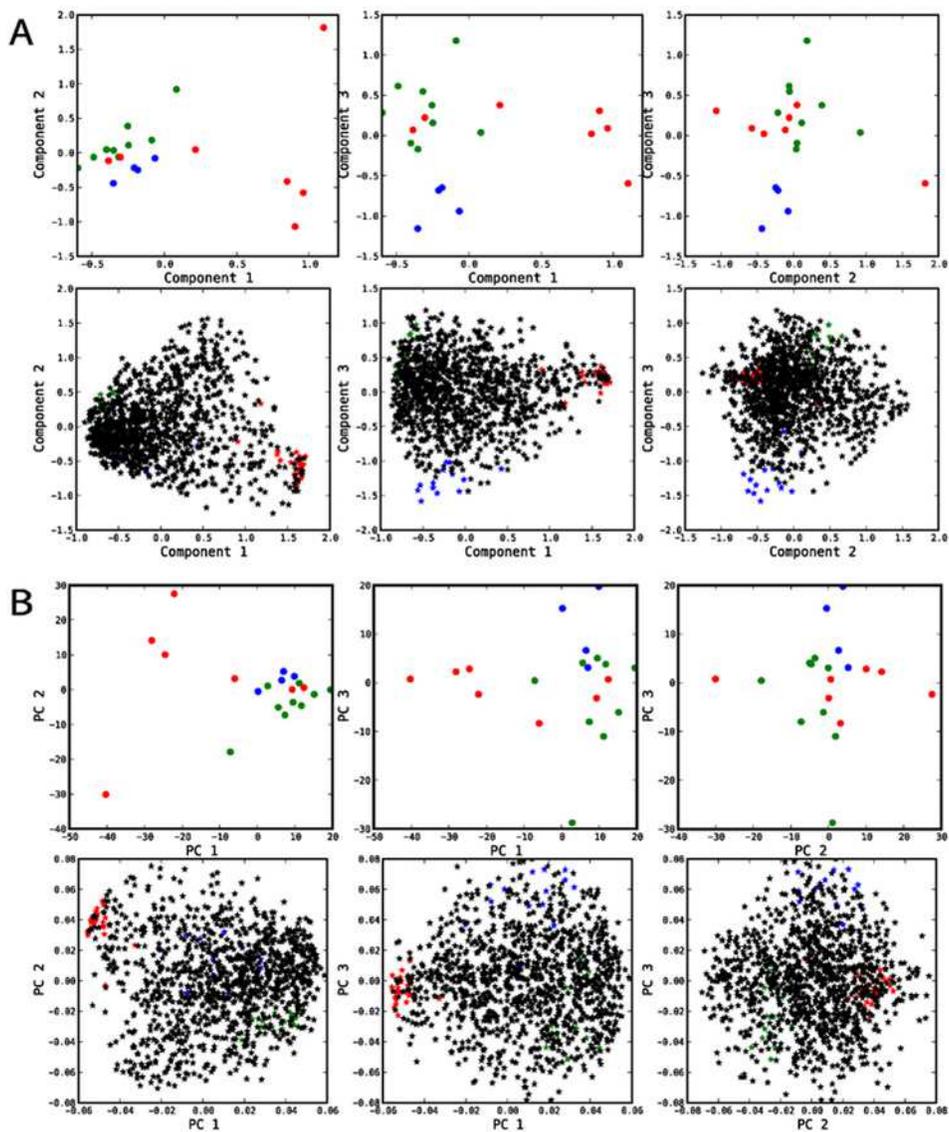}

\caption{Low-dimensional representation of samples and variables from
the microRNA data set, obtained by CUMBIA (panel \textup{A}) and PCA (panel \textup{B}).
Red markers represent samples from the CML subgroup, blue markers
correspond to the AML group and green markers to the B-ALL group.
Variables shown in one of these colors are significantly higher
expressed in the corresponding sample group than in the other two
(Student's $t$-test, one-tailed $p < 0.0005$, note that this information
was not used to obtain the visualizations, but is merely displayed to
facilitate the interpretation).}
\label{miRNAfigure}
\end{figure}

In the previous examples we have shown that for data sets where the
main nonrandom variation is attributable to small groups of samples
sharing extreme values for small groups of variables, CUMBIA can
produce sample and variable visualizations that are more informative
than those resulting from PCA and the applied projection pursuit
algorithm. Now, we consider a data set containing measurements of 1,145
microRNAs in 20 human leukemia cell lines (unpublished data). The cell
lines correspond to three different leukemia types; CML (chronic
myeloid leukemia), AML (acute myeloid leukemia) and B-ALL (B-cell acute
lymphoblastic leukemia). Figure~\ref{miRNAfigure} shows the
visualizations obtained by CUMBIA and PCA. In this case, the feature
distinguishing three of the CML samples (red markers) from the rest of
the samples contains enough variance to be picked up by PCA. The
discrimination of these samples is apparent also with CUMBIA, where
furthermore the third component effectively discriminates the AML group
(blue) from the B-ALL group (green). This effect is more readily
visible than in the PCA visualization. The CML group is biologically
heterogeneous which can also be seen in the visualizations. To
facilitate the interpretation of the visualizations, we have colored
all variables which are significantly higher expressed in one sample
group than in the others. The heterogeneity of the CML group is
reflected also here, in that some of the variables which are closely
related to the three deviating CML samples are not significantly
differentially expressed in the whole CML group. On the other hand, it
is clear that the variables which have the most negative values on the
third CUMBIA component are all highly expressed in the closely located
AML samples (blue). Scree plots for CUMBIA and PCA are available in the
Supplementary Material. We used $d=5$ principal components in the
whitening for projection pursuit, and the resulting components are
shown in the Supplementary Figures. In this case, the sample
representations from projection pursuit results are not very different
from those of CUMBIA, but the coupling between the salient sample
groups and the corresponding discriminating variables is stronger with
CUMBIA. In the absence of external annotations, this possibly enables
formulating sharper and more correct hypotheses. Applying SAMBA to this
data set returns 16 biclusters. Generally, from these biclusters it is
difficult to extract information distinguishing the three leukemia subtypes.

Taken together, the examples indicate that CUMBIA is a useful
complement to existing visualization methods in different contexts. It
can find features commonly detected by existing methods such as PCA and
projection pursuit, but also features that are difficult to find with
these methods.

\section{\texorpdfstring{Discussion and conclusions.}{Discussion and conclusions}}
We have described CUMBIA: an unsupervised algorithm for exploratory
analysis and simultaneous visualization of the samples and variables of
a multivariate data set. The basis of the algorithm is classical
multidimensional scaling (MDS), which is applied to a joint
dissimilarity matrix and produces a common low-dimensional
representation of samples and variables. The dissimilarity between a
sample and a variable is based on the expression level of the variable
in the \mbox{sample}; a~higher expression level gives a lower dissimilarity.
The dissimilarity between two samples (or two variables) is then
defined by graph distances, influenced mainly by the variables
(samples) with a high total expression level in the two samples (variables).
By applying the method to a synthetic as well as real-world data sets,
we have shown its ability to extract relevant sample and variable
groups. Compared to PCA, which is commonly used for visualization of
high-dimensional data, the proposed method is advantageous for
extracting small related variable and sample subgroups. According to
the proposed dissimilarity measure, two samples will be considered
close if they share a high value of one or a small group of variables.
This is in contrast to PCA, where the entire variable profiles are used
to calculate the distance between a pair of samples. We believe that
the proposed method may be a valuable complement to existing methods
for exploratory analysis of multivariate data, to extract closely
related sample clusters and immediately find the variables which are
responsible for the discrimination. This group of variables can then be
analyzed further and may constitute potential biomarkers for the
corresponding sample group. As described in this paper, the proposed
algorithm is mainly directed toward finding groups of samples sharing a
high expression value of a, possibly small, group of variables, but can
be adjusted to emphasize also jointly underexpressed variables.

By choosing different values of $K$ (the number of paths to average
over in the calculation of the CUMBIA dissimilarities), it is possible
to detect different structures. A small value of $K$ makes it possible
to find very small sample and variable groups but makes the method
sensitive to noisy data. With increasing $K$ the method becomes more
robust, but it is also more difficult to detect the smallest groups. In
an exploratory study, CUMBIA could be applied with different values of
$K$ to find as many potentially relevant patterns as possible.

Putting the negative eigenvalues to zero in the classical MDS as we
have done in this paper potentially discards interesting information,
as discussed by \citet{LaubMuller04}. Interestingly, in the examples
that we have given most eigenvalues are positive, but there is one
large negative eigenvalue which corresponds to an eigenvector
separating the sample objects from the variable objects. However, since
we are mainly interested in the interaction between samples and
variables, we focus on the largest positive eigenvalues of the inner
product matrix and the corresponding eigenvectors.

The induced dissimilarities from CUMBIA may be potentially useful for
clustering of samples and/or variables, for example, by hierarchical
clustering [\citet{Sneath57}; \citet{Hastieetal09}]. One would then expect
small sample groups, characterized by few variables, to be clustered
more closely than with hierarchical clustering based on, for example,
Euclidean distance. The dissimilarities can potentially also be used
for simultaneous feature and sample selection from the data set by
backward feature elimination, in a manner similar, for example, to the
``gene shaving'' [\citet{Hastieetal00}] and ``recursive feature
elimination'' [\citet{Guyonetal02}] procedures. This could be done in
the following way. First, the joint CUMBIA dissimilarity matrix for the
entire data set is calculated. Then, for each object (sample or
variable), the mean value of the $K_0$ smallest dissimilarities between
the object and all objects of the same type (i.e., samples or
variables) are calculated for a suitable choice of $K_0$. A given
fraction of the objects, consisting of those with the largest value of
the mean dissimilarity score, can then be removed. This gives a new
data matrix, with fewer samples and variables, to which the process may
be applied. This algorithm provides a sequence of nested
sample--variable biclusters. The optimal cluster size should be
determined based on a suitably chosen optimality criterion.
Furthermore, when a bicluster has been found, the included variables
and samples may be removed from the data set and another, disjoint
bicluster may be found from the resulting matrix.

\begin{supplement}
\stitle{Supplementary material}
\slink[doi]{10.1214/11-AOAS460SUPPA} 
\slink[url]{http://lib.stat.cmu.edu/aoas/460/supplementA.pdf}
\sdatatype{.pdf}
\sdescription{In the supplementary material we give a small schematic example showing the
different steps of CUMBIA. Further, we show how to emphasize both over- and
underexpressed variables in the visualization and how the choice of $K$ and $s$
affect the resulting visualization. We also provide scree plots obtained by
CUMBIA and PCA for the three data sets studied in the paper.}
\end{supplement}
\begin{supplement}
\stitle{Supplementary figures---Projection pursuit results\\}
\slink[doi,text={10.1214/11-AOAS460SUPPB}]{10.1214/11-AOAS460SUPPB} 
\slink[url]{http://lib.stat.cmu.edu/aoas/460/supplementB.pdf}
\sdatatype{.pdf}
\sdescription{The supplementary figures show the result of the
FastICA projection pursuit algorithm applied to the three data
sets considered in the paper. Note that to facilitate the interpretation
of the figures, the axes are ungraded and only the origin is marked.}
\end{supplement}


\printaddresses

\end{document}